%% file: paper.tex
\documentclass[aps,prl,twocolumn,showpacs,superscriptaddress,groupedaddress]{revtex4}  
\usepackage{graphicx}  
\usepackage{dcolumn}   
\usepackage{bbm}        
\usepackage{amssymb}   
\usepackage{amsmath}

\input{macro}

\begin{document}

\widetext


\title{Hearing the Maximum Entropy Potential of neuronal networks}
\input author_list.tex       
\date{\today}

\begin{abstract}
We consider a spike-generating stationary Markov process whose transition probabilities are known. We show that there is a canonical potential whose Gibbs distribution, obtained from the Maximum Entropy Principle (MaxEnt), is the equilibrium distribution of this process. We provide a method to compute explicitly and exactly this potential as a linear combination of \textit{spatio-temporal} interactions. In particular, our results establish an explicit relation between Maximum Entropy models and neuro-mimetic models used in spike train statistics.
\end{abstract}

\maketitle

\begin{figure}[h]
\begin{center}
\scalebox{0.4}
{
\includegraphics{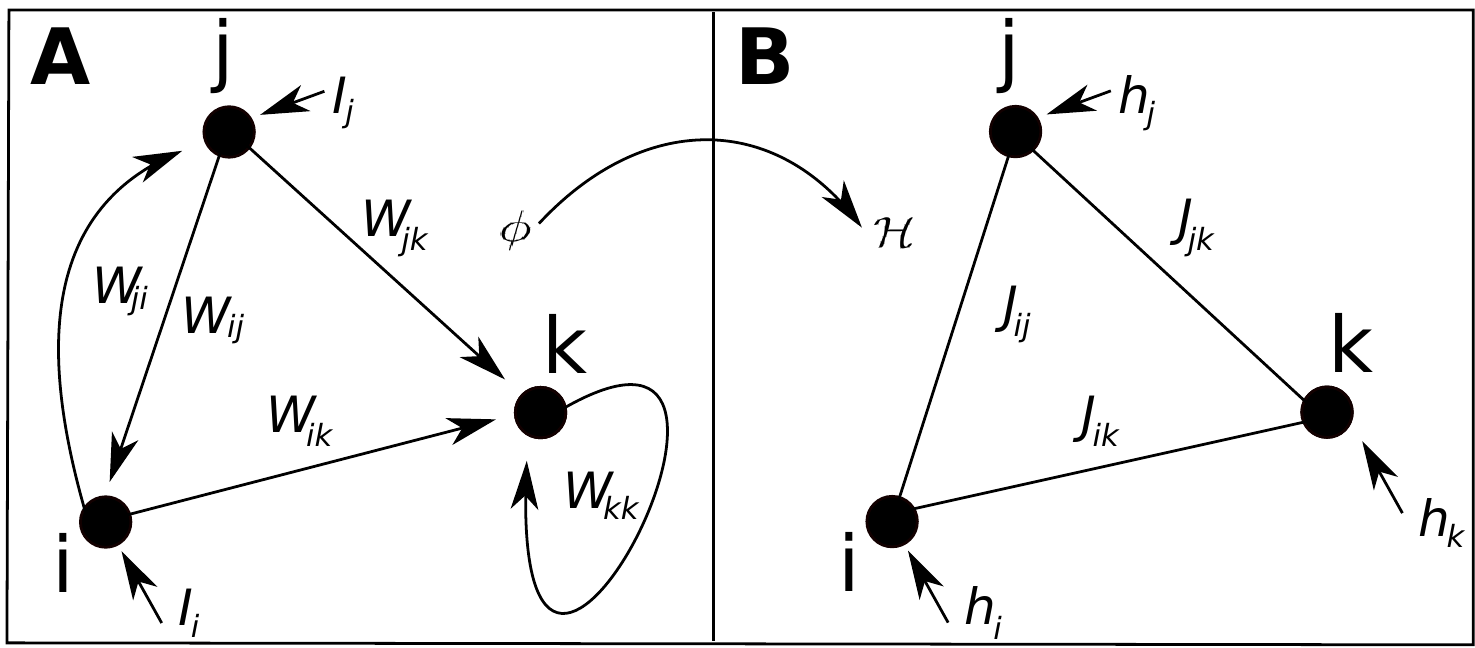}}
\caption{\label{fig:dual} \footnotesize{a. (left). Neuro-mimetic approach. Neurons are interacting via spikes synaptic weights $W_{ij}$ and stimulus $I$.
Transition probabilities are explicit functions of these parameters. b. MaxEnt statistical approach. The graph represents Ising model where only local fields and pairwise interactions are drawn. More general interactions are considered in the text. In Ising model pairwise interactions are symmetric (represented without arrows). We are looking for an explicit and exact correspondence between these two representations.}}
\end{center}
 \vspace{-15pt}
\end{figure}

The spike train response of a neuronal network to external stimuli is largely conditioned by the stimulus itself, synaptic interactions and  neuronal network history \cite{gerstner-kistler:02b}.
Understanding these dependences is a current challenge in neuroscience \cite{rieke-warland-etal:97}.
One current research
trend is based on the assumption that spikes are generated by a Markov process where the form of transition probabilities
is derived from our knowledge about neuronal network dynamics (``neuro-mimetic" models). Prominent examples are the Linear-Non Linear model (LN)  and
the Generalized Linear Model (GLM) \cite{brillinger:88,chichilnisky:01} or  Integrate-and-Fire models \cite{gerstner-kistler:02b}, \cite{cofre-cessac:13}. In all these examples the transition probabilities are explicit functions of ``structural" parameters in the neural network (synaptic weights  $\mathcal{W}$ matrix, and stimulus $\mathcal{I}$) (Fig. \ref{fig:dual}a). Another trend is based on the Maximum Entropy Principle (MaxEnt) \cite{jaynes:57}.
It consists of fixing a set of constraints,
determined as the empirical average of observables measured from the spiking activity. Maximizing the statistical entropy given those constraints provides a unique
probability, called a Gibbs distribution.
The choice of constraints determines a ``model". Prominent examples are the Ising model \cite{schneidman-berry-etal:06,shlens-field-etal:06} where constraints are firing rates and probabilities that $2$ neurons fire at the same time, the Ganmor-Schneidman-Segev model \cite{ganmor-segev-etal:11a}, which considers additionally the probability of triplets and quadruplets of spikes, or the Tka\v{c}ik et al model \cite{tkacik-marre-etal:13} where the probability that $K$ out of $N$ cells in the network generate simultaneous action potentials is an additional constraint. In these examples dynamics has no memory; but Markovian models where the probability of a spike pattern at a given time depends on the spike history can be considered as well \cite{marre-boustani-etal:09,vasquez-marre-etal:12}.     
These models depends on a set of parameters (Lagrange multipliers) which are naturally interpreted, in the case of pairwise constraints, as ``functional interactions" between neurons \cite{ganmor-segev-etal:11a} (see fig \ref{fig:dual}b for the Ising model). 

To summarize, (at least) two different representations can be used to analyze spike train statistics in neuronal networks (fig. \ref{fig:dual}).
The goal of this paper is to establish an explicit and exact link between
these two representations. An important work in this direction has been done in \cite{cocco-leibler-etal:09}, linking
the ``neuro-mimetic" Integrate and Fire model to Ising model. We propose here a generalization which allows us to handle
more general types of neuro-mimetic models as well as general MaxEnt distributions (including memory). The method
we used is based on Hammersley-Clifford decomposition theorem \cite{hammersley-clifford:71} and periodic orbits decomposition 
in ergodic theory \cite{pollicott-weiss:03}. The techniques are therefore different from \cite{cocco-leibler-etal:09}.
More generally, we solve the following problem.  Assuming that a spike
train has been generated by a Markov process where transition probabilities are given (and strictly positive), can we construct
a MaxEnt model, with a minimum of constraints, reproducing exactly the (spatio-temporal) statistics of this
process? When the Markov process is generated by a neuro-mimetic model, this establishes an exact correspondence between the structural parameters and the parameters of the MaxEnt.

We study a network composed of $N$ neurons.
Time has been discretized so that a neuron can at most fire
a spike within one time bin. A spike train $\omega$ is represented by a binary
time series with entries $\omega_k(n)=1$ if  neuron $k$ fires at time $n$ and $\omega_k(n)=0$ otherwise. 
The \textit{spiking pattern} at time $n$ is the 
  vector  $\omega(n) = \bra{\omega_k(n)}_{k=1}^{N}$.
A {\em spike block} $\bloc{n_1}{n_2}$ is an ordered list of spiking patterns
 where spike times range from $n_1$ to $n_2$. 
 
In a neuronal network the probability that the spike pattern $\omega(n)$ occurs at time $n$  
is the result of the complex membrane potentials dynamics \cite{gerstner-kistler:02b}. A simplification consists
of assuming that this probability is only a function of the spike history up to a certain memory
depth $D$. This provides a family of conditional probabilities $\Probct{n}{\omega(n)}{\bloc{n-D}{n-1}}$ which
may depend explicitly on time as indicated by the sub-index $n$. They define a Markov chain. In neuro-mimetic models
these probabilities depends on parameters that mimics biophysical quantities such as  synaptic weights matrix $\cW$ and stimulus
$\mathcal{I}$ e.g.  :
\beq\label{GLM}
\Probct{n}{\omega(n)}{\bloc{n-D}{n-1}}= g\bra{\omega(n),\cW X(\bloc{n-D}{n-1}) + i(\bloc{n-D}{n})},
\eeq
 where $X(\bloc{n-D}{n-1})$ integrates the spike history, whereas $i(\bloc{n-D}{n})$ integrates the stimulus effect.
$g$ is a non linear function (typically sigmoid) of its second argument. 
The probability depends also on the spike pattern at time $n$ (first argument of $g$).

In this paper, we make two assumptions. (i) Transition probabilities do not depend on time i.e.  spike statistics is stationary. We can then drop the index $n$ in $\Probct{n}{\omega(D)}{\bloc{0}{D-1}}$; (ii) $\Probc{\omega(D)}{\bloc{0}{D-1}} > 0$.
This ensures that there is a unique invariant distribution for the Markov chain, denoted by $\mu$.  We note:
\beq\label{Phi}
\phi\pare{\bloc{0}{D}} = \log \Probc{\omega(D)}{\bloc{0}{D-1}}.
\eeq
From the Chapman-Kolmogorov equation we have,
for $n_2 -n_1+1 > D$, $\moy{\bloc{n_1}{n_2}} = e^{\sum_{n=n_1}^{n_2-D} \phi(\bloc{n}{n+D})} \moy{\bloc{n_1}{n_1+D-1}}$.
This suggests that $\mu$ is a Gibbs distribution with potential $\phi$ and normalization factor $1$. 
We see below that this is indeed the case. We call $\phi$ a \textit{normalized potential}.  

The MaxEnt  also defines a Markov chain in the following way. A potential of range $R=D+1$ is a function $\H$ which associates
to a spike block $\bloc{0}{D}$ a real value. We assume $\H(\bloc{0}{D}) > - \infty$. Any such potential can be written:
\beq\label{Hh}
\H(\bloc{0}{D}) \, =\sum_{l=0}^{{L}} h_l \, m_l(\bloc{0}{D}),
\eeq
where $L=2^{NR}-1$. $h_l$'s are real numbers whereas the function $m_l$ with $m_l(\bloc{0}{D})=\prod_{u=1}^r \omega_{k_u}(n_u)$ is called a \textit{monomial}. $r$ is the \textit{degree} of the monomial. By analogy with spin systems, we see from (\ref{Hh}) that monomials somewhat constitute \textit{spatio-temporal} interactions: degree $1$ monomials corresponds to ``self-interactions", degree $2$ to pairwise interactions, and so on.  
In many examples most $h_l$'s are equal to zero. For example, Ising model considers
only monomials of the form $\omega_i(0)$ (singlets) or $\omega_i(0)\omega_j(0)$ (pairwise events). 
More generally, the potential (\ref{Hh}) considers spatio-temporal events occurring within a time horizon $R$. 

There is a unique stationary probability $\mu$, called \textit{Gibbs distribution with potential $\H$} satisfying:  
\beq\label{VarPrinc}
\p{\H}=\sup_{\nu \in \cM} \pare{\s{\nu} \, + \, \noy{\H}}=
\s{\mu} \, + \, \moy{\H},
\eeq
where $\cM$ is the set of stationary probabilities on the set of spike trains, whereas $\s{\nu} \, = \, - \, \lim_{n \to \infty} \frac{1}{n+1} \, \sum_{\bloc{0}{n}} \, \noy{\bloc{0}{n}} \, \log \noy{\bloc{0}{n}}$ is the entropy of $\nu$. $\noy{\H}=\sum_{l=0}^L h_l \noy{m_l}$ is the average of $\H$ with respect to $\nu$. The average value of each $m_l$, $\noy{m_l}$, is fixed e.g. by experiments. This  constitutes a set of constraints in (\ref{VarPrinc}).
The quantity $\p{\H}$ is the \textit{free energy}; this is a convex function of $h_l$'s and $\frac{\partial \p{\H}}{\partial h_{l}} = \moy{m_{l}}$. 

Two distinct potentials $\H^{(1)},\H^{(2)}$ of range $R$ can correspond to the same Gibbs distribution (we call them  \textit{equivalent}). A standard result in ergodic theory states that  $\H^{(1)}$ and $\H^{(2)}$
are equivalent if and only if 
there exists a range $D$ function $f$ such that \cite{bowen:08}:
\begin{equation}\label{Cohomology}
\H^{(2)}\pare{\bloc{0}{D}}=\H^{(1)}\pare{\bloc{0}{D}} - f\pare{\bloc{0}{D-1}}+ f\pare{\bloc{1}{D}} + \Delta,
\end{equation} 
where $\Delta =\p{\H^{(2)}} - \p{\H^{(1)}}$. This relation establishes a \textit{strict} equivalence 
 and does not correspond e.g. to renormalization. The ``if" part is easy. Indeed, injecting $\H^{(2)}$ in the variational form (\ref{VarPrinc}) the terms corresponding to $f$ cancels
 because $\nu$ is time-translation invariant. Therefore, the supremum is reached for
 the same Gibbs distribution as $\H^{(1)}$ whereas $\Delta$ is indeed the difference of free energies. The ``only if" part is more tricky.
 
Let us give a key example. To any potential $\H$ of the form (\ref{Hh}) corresponds a unique equivalent \textit{normalized} potential (\ref{Phi}).
In this case the function
$\cG(\bloc{0}{D}) =  f\pare{\bloc{0}{D-1}} - f\pare{\bloc{1}{D}} - \Delta$ acts as normalization function. For potentials of range $R=1$ ($D=0$), $\cG = \log Z$, where $Z$ is the partition function and one obtains the standard normalization of spatial Gibbs distributions. In the spatio-temporal case $\cG$ is written in terms of the largest eigenvalue and the corresponding right eigenvector of a transfer matrix associated with $\H$ \cite{georgii:88}.

This result establishes a relation between Markov chain-normalized potentials (\ref{Phi})
on one hand and potentials of the form (\ref{Hh}) on the other hand (the arrow $\phi   \rightarrow \H$ in fig. \ref{fig:dual}). However, due to the arbitrariness in the choice of $f$ in (\ref{Cohomology}) there are infinitely many potentials
$\H$ corresponding to the same Gibbs distribution (the same normalized potential $\phi)$.
So the next question is: Can we find a canonical form of $\H$ with a minimal number of terms
 equivalent to a given normalized potential ? The situation is a bit like normal forms in bifurcations theory where variable changes permits to eliminate non resonant terms in the Taylor expansion of the vector field \cite{arnold:83}. Here, the role of the variable changes is played by $f$. By suitable choices of $f$ one should be able to eliminate some monomials in the expansion (\ref{Hh}). An evident situation corresponds to monomials related by time translation, e.g. $\omega_i(0)$ and $\omega_i(1)$: since any $\nu \in \cM$ is time translation invariant
 $\noy{\omega_i(0)}=\noy{\omega_i(1)}$, the firing rate of neuron $i$. Such monomials correspond to the same constraint in (\ref{VarPrinc}) and can therefore be eliminated.  A potential where monomials, related by time translation to a given monomial, have been eliminated (the corresponding $h_l$ vanishes) is called \textit{canonical}. A canonical potential contains thus, in general, $2^{NR} - 2^{N(R-1)}$ terms. One can show that \textit{only} these monomials can be eliminated by a transformation like (\ref{Cohomology}). This implies that two canonical potentials are equivalent \textit{if and only if} their coefficients $h_l$, $l>0$, are equal \cite{cessac-cofre:13c}. There is still an arbitrariness due to the term $h_0$ (``Gauge" invariance). One can set it equal $0$ without loss of generality. In this way, there is only one canonical potential, with a minimal number of monomials, corresponding to a given stationary Markov chain.  

The next question is: how to compute the coefficients of the canonical potential $\H$ from the knowledge of $\phi$ ? Given a spike block $\om{l_1}$, a \textit{periodic orbit} of period $\tau$ is a sequence of spike blocks $\om{l_n}$, $n=1 \dots $ where $\om{l_{k \tau +  n}} = \om{l_n}$, $k \geq 0$, $0 \leq n \leq \tau$. Now, from (\ref{Cohomology}) we have, for such a periodic orbit, 
$\sum_{n=1}^\tau \H^{(2)}\pare{\om{l_n}}= \sum_{n=1}^\tau \H^{(1)}\pare{\om{l_n}} + \tau \Delta$, because the $f$-terms disappear when summed along a periodic orbit. It follows that
the sum of a potential along a periodic orbit is an invariant (up to the constant term $\tau \Delta$) in the class of equivalent potentials. This is a classical result in ergodic theory extending to infinite range potentials \cite{livsic:72}. Therefore, in the correspondence between $\phi$ and its canonical potential $\H$ we have:
\beq \label{Livsic}
\sum_{n=1}^\tau \sum_{l=0}^{{L}} h_l  m_l(\om{l_n})= \sum_{n=1}^\tau \phi \pare{\om{l_n}}  + \tau \p{\H}
\eeq 

This equation is especially useful if one takes advantage of an existing hierarchy between blocks and between monomials, the Hammersley-Clifford hierarchy \cite{hammersley-clifford:71}. Each spike block is associated to a unique integer (index) $l =\sum_{k=1}^N \sum_{n=0}^D 2^{n\,N+k-1} \, \omega_k(n)$. We denote $\om{l}$ the spike block corresponding to the index $l$. Likewise, since a monomial is defined by a set of spike events $(k_u, n_u)$, one can associate to each monomial a spike block or ``mask" where the only bits '$1$' are located at $(k_u,n_u)$, $u=1, \dots, r$. This mask has therefore an index.
Whereas the labeling of monomials in (\ref{Hh}) was arbitrary, $m_l$ denotes from
now on the monomial with mask $\om{l}$. We define the block inclusion $\sqsubseteq$ by
$\bloc{0}{D} \sqsubseteq \seq{\omega'}{0}{D}$ if $\omega_k(n)=1 \Rightarrow \omega'_k(n)=1$, with the convention
that the block of degree $0$, $\om{0}$, is included in all blocks. This inclusion defines the Hammersley-Clifford hierarchy. 
Then, for two integers $l,l'$, $m_{l'}(\om{l})=1$ if and only if $\om{l'}  \sqsubseteq \om{l}$.
It follows that (\ref{Livsic}) becomes 
\beq \label{Livsic2}
\sum_{n=1}^\tau \sum_{l \sqsubseteq l_n} h_l  \, m_l(\om{l_n})= \sum_{n=1}^\tau \phi (\om{l_n})  + \tau \p{\H}
\eeq%
where, with a slight abuse of notations $l \sqsubseteq l_n$ stands for $\om{l}  \sqsubseteq \om{l_n}$.

What is the use of this relation ? Start from the first mask in hierarchy, the mask $\om{0}$ containing only $0$'s, whose corresponding monomial is $m_0=1$
and consider the periodic orbit, of period $1$, $\Set{\om{0}}$. The application of (\ref{Livsic2}) gives $h_0 = \phi (\om{0}) + \p{\H}$ and since we choose $h_0=0$ for the canonical potential we obtain a direct way to compute the free energy $\p{\H} = - \phi (\om{0})$. In the memory-less case one has $\p{\H} = \log Z$. To save space we now give the following examples with $N=2,R=2$ (the generalization is straightforward). Consider the mask
$\om{l_1}=\mbloc{0 & 1\\0 & 0}$, corresponding to the monomial $\omega_1(1)$, and the periodic orbit
obtained by a $R$-circular shift of this block: $\Set{\om{l_1}=\mbloc{0 & 1\\0 & 0}, \om{l_2}\mbloc{1 & 0\\0 & 0}}$. Since $\om{l_1}$ and $\om{l_2}$ are related by time translation the coefficient of one of these monomials is set to $0$ in the canonical potential $\H$. We use the convention to keep the monomial $m_{l_1}$ whose mask contains a $1$ in the last column. This convention extends to the monomials considered below. Then, from
 (\ref{Livsic2}) we obtain $h_{l_1}= \phi (\om{l_1}) +
\phi (\om{l_2}) + 2\p{\H}$. This procedure gives all degree one interactions terms. 

The idea
is then to proceed recursively, by increasing degree in the Hammersley-Clifford hierarchy. Let us consider pairwise interactions.  For Ising coefficients, corresponding
to blocks of the form $\om{l_1}=\mbloc{0 & 1\\0 & 1}$, the procedure is the same as above:
take the periodic orbit $\Set{\om{l_1}=\mbloc{0 & 1\\0 & 1}, \om{l_2}=\mbloc{1 & 0\\1 & 0}}$,
remark that $\om{l_2}$ corresponds to a non canonical interaction, and compute the pairwise
coefficient of $m_{l_1}$ where the l.h.s. of (\ref{Livsic2}) contains $h_{l_1}$ and coefficients corresponding to masks of lower degree which have been already computed. Then the Ising coefficient  $h_{l_1}$ can be determined (see eq. (\ref{Jijising}) for an explicit form).
For the one step of memory pairwise coefficients (e.g. $\omega_1(0) \omega_2(1)$) appearing in \cite{marre-boustani-etal:09}
the situation is slightly different. Consider e.g. the mask  $\om{l_1}=\mbloc{1 & 0\\0 & 1}$
corresponding to the monomial $\omega_1(0) \omega_2(1)$. The periodic orbit $\Set{\mbloc{1 & 0\\0 & 1},\mbloc{0 & 1\\1 & 0}}$ is not useful because it leads to $2$ unknown in eq. (\ref{Livsic2}). Instead, the orbit $\Set{\mbloc{1 & 0\\0 & 1},\mbloc{0 & 0\\1 & 0},\mbloc{0 & 0\\0 & 0}, \mbloc{0 & 1\\0 & 0}}$, of period $4$ can be used. When applying (\ref{Livsic2}) there is only one unknown, corresponding to the mask $\mbloc{1 & 0\\0 & 1}$ whereas the coefficients of the other blocks have a lower degree and have therefore been already computed.
The general procedure for arbitrary blocks is described in \cite{cessac-cofre:13c}.

Obviously, when getting to larger degrees the method becomes rapidly intractable because of the exponential increase in the number of terms. The hope is that the influence of monomials decays rapidly with their  degree. Additionally, applying it to real data where transition probabilities are not exactly known leads to severe difficulties.  These aspects will be treated in a separated paper. Our goal here was to answer the questions asked in the introduction and, as a by-product, to establish a link between neuro-mimetic models for spike train statistics 
and MaxEnt models. This goal is now achieved. To illustrate this we show how the coefficients shaping the Ising model are written in terms of transition probabilities of a Markov process with range $R>1$. The Ising potential only provides an approximation of the equilibrium distribution of the process, corresponding to degree $2$ expansion in the Hammersley-Clifford hierarchy, where, besides, memory effects are neglected.
Let us write the Ising potential in the form $\sum_{i=1}^N \h_i \omega_i(0) + \sum_{i<j} \J_{ij} \omega_i(0) \omega_j(0)$. 
We keep here the spike description with $0$'s and $1$'s but the relation with the classical
Ising Hamiltonian with spins $\in \Set{-1,1}$ is straightforward.
From the method described above we obtain:
\beq \label{h_i}
\h_i =  \sum_{n=1}^{R} \phi\pare{\om{\sigma^n l_i}} -R\phi(\om{0}),
\eeq
where $\om{l_i}$ is the mask  having $0$'s everywhere but on  row $i$-column $D$ whereas
$\sigma$ denotes the periodic shift. 
In the same way,  denoting $\om{l_{ij}}$ the mask with $0$'s everywhere but on rows $i,j$-column $D$:
\beq\label{Jijising}
\J_{ij}=
\sum_{n=1}^{R} \phi\pare{\om{\sigma^n l_{ij}}} 
- \h_i - \h_j 
-R \phi(\om{0}).
\eeq
Note that these coefficients depend on the memory depth of the Markov process.

When $\phi$ is derived from a neuro-mimetic model (e.g. eq. (\ref{GLM})), it follows that the ``local field"  $\h_i$ depends non linearly on the complete stimulus $\mathcal{I}$ (not only the stimulus applied to neuron $i$); this is also  a non linear function of the synaptic weights matrix $\cW$. This is not that surprising. Even in an Ising model of two
neurons with no memory, a strong favorable pairwise
interaction between the two neurons firing simultaneously
will increase the average firing rate of both neurons, even in the
absence of an external field. 
Likewise, $\J_{ij}$ depends on the whole  synaptic weights matrix $\cW$  and not only on the connection between $i$ and $j$. This example clearly shows that there is no straightforward relation
between the so-called ``functional connectivity" in Ising model ($\J_{ij}$'s) and the real connectivity.

It results from our analysis that the $h_l$'s of a canonical potential corresponding to a neuro-mimetic model are \textit{generically} non zero: considering e.g. \textit{random} synaptic weights $W_{ij}$, the probability that some $h_l$ in (\ref{Livsic2}) vanishes is zero. 
Therefore, while there are $O(N^2)$ parameters to constraint the transition probabilities in neuro-mimetic models, the number of MaxEnt parameters increases exponentially fast with $N$. Thus, there is a great amount of redundant information in the $h_l$ which are related by non linear relations.
However, real neural networks are non generic: synaptic weights are not drawn at random but result from a long phylogenetic and ontogenetic evolution. When trying to ``explain" spike statistics of real neural networks with the Maximum Entropy Principle, one is seeking some general laws that has to be expressed with relatively few phenomenological parameters in the potential (\ref{Hh}). The hope is that many coefficients coming from real data are $0$ or close to $0$. This could explain the efficiency of pairwise  MaxEnt models \cite{bialek-ranganathan:07} for spike trains analysis. Our method  provides a way do detect this, if the l.h.s. in (\ref{Livsic2}) is close to $0$ \cite{cessac-cofre:13c}.

More generally our method opens up new possibilities which allow a better understanding of the role of different neural network topologies and stimulus on spike responses. It is not limited to spike trains however and could also impact different areas of scientific knowledge where binary time series are considered.

\input{acknowledgements}

\bibliographystyle{plain}
\bibliography{../odyssee,../biblio}

\end{document}

%% file: macro.tex
\newcommand\seq[3]{{#1}_{#2}^{#3}}
\newcommand{\bd}{\begin{document}}
\newcommand{\ed}{\end{document}}
\newcommand{\beq}{\begin{equation}}
\newcommand{\eeq}{\end{equation}}
\newcommand{\bef}{\begin{figure}}
\newcommand{\enf}{\end{figure}}
\newcommand{\bea}{\begin{eqnarray}}
\newcommand{\eea}{\end{eqnarray}}
\newcommand{\baR}{\begin{array}}
\newcommand{\eaR}{\end{array}}
\newcommand{\bc}{\begin{center}}
\newcommand{\ec}{\end{center}}
\newcommand{\ben}{\begin{enumerate}}
\newcommand{\een}{\end{enumerate}}
\newcommand{\bit}{\begin{itemize}}
\newcommand{\eit}{\end{itemize}}
\newcommand{\su}{\section}
\newcommand{\ssu}{\subsection}
\newcommand{\sssu}{\subsubsection}
\newcommand\ssssu[1]{\textbf{#1}}
\newcommand{\nid}{\noindent}
\newcommand{\nnb}{\nonumber}
\newcommand\bloc[2]{{\omega}_{#1}^{#2}}
\newcommand\blocp[2]{{\omega'}_{#1}^{#2}}
\newcommand{\un}{\mathbbm{1}}
\newcommand{\setZ}{\mathbbm{Z}}
\newcommand{\setN}{\mathbbm{N}}
\newcommand{\setR}{\mathbbm{R}}
\newcommand{\pare}[1]{\left(\, #1 \, \right)}
\newcommand{\scal}[2]{\langle\, #1 | #2 \, \rangle}
\newcommand{\bra}[1]{\left[\, #1 \, \right]}
\newcommand{\brac}[2]{\left[\, #1 \, | \, #2 \right]}
\newcommand{\Set}[1]{\left\{\, #1 \, \right\}}
\newcommand{\Setc}[2]{\left\{\, #1 \, ; \, #2 \right\}}
\newcommand{\Prob}[1]{P\left[\, #1 \, \right]}
\newcommand{\Probt}[2]{P_{#1}\left[\, #2 \, \right]}
\newcommand{\PT}[1]{\pi^{(T)}\left[\, #1 \, \right]}
\newcommand{\PTo}[1]{\pi^{(T)}_\omega\left[\, #1 \, \right]}
\newcommand{\pT}{\pi^{(T)}}
\newcommand{\Probex}[1]{P^{(T)}\left[\, #1 \, \right]}
\newcommand{\probc}[2]{P[#1 \, \left| \, #2 \right.]}
\newcommand{\Probc}[2]{P\bra{#1 \, \left| \, #2 \right.}}
\newcommand{\Probct}[3]{P_{#1}\bra{#2 \, \left| \, #3 \right.}}
\newcommand{\Probcex}[2]{P^{(T)}\bra{#1 \, \left| \, #2 \right.}}
\newcommand{\Probcp}[2]{P^{(ap)}\bra{#1 \, \left| \, #2 \right.}}
\newcommand{\abs}[1]{\left|\, #1 \, \right|}
\newcommand{\deq}{\stackrel {\rm def}{=}}
\newcommand{\sqsubsetneq}{\stackrel {\tiny\sqsubset}{\tiny \neq}}
\newcommand\moy[1]{\mu\bra{#1}}
\newcommand\noy[1]{\nu\bra{#1}}
\newcommand\cB{{\mathcal{B}}}
\newcommand\cC{{\mathcal{C}}}
\newcommand\cD{{\mathcal{D}}}
\newcommand\cT{{\mathcal{T}}}
\renewcommand\H{{\cH}}
\newcommand{\h}{\mathtt{h}}
\newcommand{\J}{\mathtt{J}}
\newcommand\s[1]{{\cS\bra{#1}}}
\newcommand\tH{{\tilde{H}}}
\renewcommand\th{{\tilde{h}}}
\newcommand\tFi{{\tilde{\Phi}}}
\newcommand\tfi{{\tilde{\phi}}}
\newcommand\tF{{\tilde{F}}}
\newcommand\tG{{\tilde{G}}}
\newcommand\tu{{\tilde{u}}}
\newcommand\tv{{\tilde{v}}}
\renewcommand{\sb}{s_\bbeta}
\newcommand\cE{{\mathcal{E}}}
\newcommand\cG{{\cal G}}
\newcommand\cH{{\mathcal{H}}}
\newcommand\cI{{\mathcal{I}}}
\newcommand\cK{{\cal K}}
\newcommand\cL{{\cal L}}
\newcommand\cM{{\mathcal{M}}}
\newcommand\cS{{\cal S}}
\newcommand\cP{{\mathcal{P}}}
\newcommand\cO{{\cal O}}
\newcommand\cU{{\cal U}}
\newcommand\cW{{\cal W}}
\newcommand\f{{\bm{f}}}
\newcommand\be{{\bm{e}}}
\newcommand\bh{{\bm{h}}}
\newcommand\bxi{{\bm{\xi}}}
\newcommand\bn{{\bm{n}}}
\newcommand\bv{{\bm{v}}}
\newcommand\bF{{\bm{F}}}
\newcommand\bo{{\bm{o}}}
\newcommand\blambda{{\bm{\lambda}}}
\newcommand\bbeta{{\bm{\beta}}}
\newcommand\gk[1]{g_k\pare{#1}}
\newcommand\gLk{g_{L,k}}
\newcommand\akj[1]{\alpha_{kj}(#1)}
\newcommand\sif[1]{\seq{\omega}{#1}{D}}
\newcommand\Xk[1]{X_k({#1})}
\newcommand\Ik[1]{I_k({#1})}
\newcommand\Skj[1]{S_{kj}({#1})}
\newcommand\X[2]{X_{#1}\pare{#2}}
\newcommand\et[2]{\eta_{#1}\pare{#2}}
\newcommand\tO[2]{\tau_{#1}\pare{#2}}
\newcommand\Vd[2]{\mathcal{V}^{(det)}_{#1}\pare{#2}}
\newcommand\XkD{\Xk{\bloc{0}{D-1}}}
\newcommand\Xkr{X_k^r}
\newcommand\tLk{\tau_{L,k}}
\newcommand\moyc[2]{\mu\bra{#1 \, | \, #2}}
\newcommand\Vkdet[1]{V_k^{(det)}({#1})}
\newcommand\Vksyn[1]{V_k^{(syn)}({#1})}
\newcommand\Vkext[1]{V_k^{(ext)}({#1})}
\newcommand\Vknoise[1]{V_k^{(noise)}({#1})}
\newcommand\sk[1]{\sigma_k({#1})}
\newcommand\skr{\sigma_k^r\pare{\bloc{0}{D-1}}}
\newcommand{\ent}[1]{\left[\, #1 \, \right]}
\newcommand\vm[1]{var_m\left[#1 \right]}
\newcommand\eg[1]{\stackrel {\rm #1}{=}}
\newcommand\tk[1]{\tau_k\pare{#1}}
\newcommand\tko{\tau_k(t,\omega)}
\newcommand\cBm[1]{\cB^{-1}\pare{#1}}
\newtheorem{theorem}{Theorem}
\newcommand{\bth}{\begin{theorem}}
\newcommand{\enth}{\end{theorem}}
\newcommand\omD[1]{\seq{\bra{\omega^{(#1)}}}{0}{D-1}}
\newcommand\omz[1]{\omega^{(#1)}(D)}
\newcommand\om[1]{\omega^{(#1)}}
\newcommand\Om[1]{\Omega^{(#1)}}
\newcommand\skjq{s_{kj;q}}
\newcommand\ikq{i_{k;q}}
\newcommand\skjqs{p_{kj;q_s}}
\newcommand\ikqs{i_{k;q_s}}
\newcommand\ikqsu{i_{k;q_{s_u}}}
\newcommand\ikqu{i_{k;q_u}}
\newcommand\xkq{x_{k;q}}
\newcommand\skjqv{s_{kj;q_{v}}}
\newcommand\tjr{t_j^{(r)}(\omega)}
\renewcommand\S[2]{S_{#1}\pare{#2}}
\newcommand{\im}{Im}
\newcommand\rpf{R}
\newcommand\rpfc[1]{R\pare{#1}}
\newcommand\lpf{L}
\newcommand\lpfc[1]{L\pare{#1}}
\newcommand{\zb}{\zeta_{\bbeta}}
\newcommand{\etab}{\eta_{\bbeta}}
\newcommand{\mex}{\mu^{\ast}}
\newcommand{\phex}{\phi^{\ast}}
\newcommand{\Phex}{\Phi^{\ast}}
\newcommand{\Hex}{\cH^{\ast}}
\newcommand{\hex}{h^{\ast}}
\newcommand{\map}{\mu^{(\chi)}}
\newcommand{\Hap}{\cH^{(\chi)}}
\newcommand{\hap}{h^{(\chi)}}
\newcommand{\siex}{\sigma^{\ast}}
\newcommand{\cn}{\cC^{\ast}}
\newcommand\p[1]{\cP\bra{#1}}
\newcommand\w[2]{w_{#1}^{(#2)}}
\newcommand\Gb{\cG_\bbeta}
\newcommand\K[2]{\cK_{\small{#1; \,  #2}}}
\renewcommand\d[2]{\delta_{\small{#1; \,  #2}}}
\newcommand\blp[3]{\omega^{#3,(#1)}_{#2}}
\newcommand\blpb[3]{\overline{\omega^{#3,(#1)}_{#2}}}
\newcommand\moyex[1]{\mu^{\ast}\bra{#1}}
\newcommand\E[1]{{\cal E}_{#1}}
\newcommand{\tc}{\tau(\cC)}
\newcommand{\notsqsupset}{\cancel{\sqsupset}}
\newcommand{\sm}{\sigma^{-1}}
\newcommand{\mbloc}[1]{\tiny{\bra{\begin{array}{ccccccc}#1\end{array}}}}

%% file: author_list.tex
\author{R.~Cofr\'{e}} \affiliation{NeuroMathComp team (INRIA, UNSA LJAD) 2004 Route des Lucioles, 06902 Sophia-Antipolis, France
}
\author{B.~Cessac} \affiliation{NeuroMathComp team (INRIA, UNSA LJAD) 2004 Route des Lucioles, 06902 Sophia-Antipolis, France
}

\vskip 0.25cm

%% file: acknowledgements.tex
%
This work was supported by the French ministry of Research and University of Nice (EDSTIC), INRIA, ERC-NERVI number 227747, KEOPS ANR-CONICYT and European Union Project $\#$ FP7-269921 (BrainScales), Renvision $\#$ 600847. We thank the reviewers for constructive criticism.